\documentstyle[12pt]{article}
\catcode`\@=11

\@addtoreset{equation}{section}
\def\theequation{\arabic{section}.\arabic{equation}}

\catcode`\@=11

\def\section{\@startsection{section}{1}{\z@}{3.5ex plus 1ex minus
   .2ex}{2.3ex plus .2ex}{\large\bf}}

\def\eqnarray{\let\@currentlabel=\theequation\refstepcounter{equation}
    \global\@eqnswtrue
    \global\@eqcnt\z@\tabskip\@centering\let\\=\@eqncr
  $$\halign to \displaywidth\bgroup\@eqnsel\hskip\@centering
   $\displaystyle\tabskip\z@{##}$&\global\@eqcnt\@ne
      \hfil${{}##{}}$\hfil
      &\global\@eqcnt\tw@ $\displaystyle\tabskip\z@{##}$\hfil
       \tabskip\@centering&\llap{##}\tabskip\z@\cr}

 \def\lefteqn#1{\hbox to 4\arraycolsep{$\displaystyle #1$\hss}}

\newcommand{\req}[1]{(\ref{#1})}
\newcommand{\be}{\begin{equation}}
\newcommand{\ee}{\end{equation}}
\newcommand{\bea}{\begin{eqnarray}}
\newcommand{\eea}{\end{eqnarray}}

\def\IR{{\hbox{{\rm I}\kern-.2em\hbox{\rm R}}}}
\def\IH{{\hbox{{\rm I}\kern-.2em\hbox{\rm H}}}}
\def\IC{{\ \hbox{{\rm I}\kern-.6em\hbox{\bf C}}}}
\def\IZ{{\hbox{{\rm Z}\kern-.4em\hbox{\rm Z}}}}
\def\ID{{\hbox{{\sl 1}\kern-.3em\hbox{\sl I}}}}

\begin{document}

\begin{titlepage}

\vspace{.5in}
\begin{flushright}
UCD-96-08\\
January 1996\\
quant-ph/9604019 \\
\end{flushright}
\vspace{.5in}

\begin{center}
{\large\bf Constrained Coherent States}\\

\vspace{.4in}
M.\  C.\  A{\sc shworth} \footnote{\it email: mikea@landau.ucdavis.edu}\\
{\small\it Department of Physics}\\
{\small\it University of California}\\
{\small\it Davis, California 95616 USA}\\
\end{center}

\vspace{.5in}
\begin{center}
{\bf Abstract}\\
\end{center}

\begin{center}
\begin{minipage}{5in}
{\small
Coherent states possess a regularized path
integral and give a natural relation between classical variables
and quantum operators.  Recent work by Klauder and Whiting has
included extended
variables, that can be thought of as gauge fields, into this formalism.
In this paper, I consider the next step, and look at the roll of
first class constraints.}

\end{minipage}
\end{center}
\end{titlepage}
\addtocounter{footnote}{-1}

\section{Introduction}

Coherent states were first introduced as non-spreading wave packets for
quantum oscillators by Schr\"odinger in the 1920's.
Later this system of states was used for many physical
applications such as quantum optics, spin waves, superfluidity, solitons,
etc.
In addition to these applications, coherent states have been used to
address more fundamental issues in quantum mechanics \cite{KlauderS}
\cite{Perel} . Klauder and others have been
developing a well-defined regularized path integral using coherent state
representations.  This formalism contains a natural relationship between
classical variables and their corresponding quantum operators.  Quantum
mechanics is also placed on a geometrical foundation.
Therefore, a preferred set of coordinates is no longer necessary to quantize
a classical system. For a good review see \cite{Klauder4}
- \cite{Klauder8}.

In a recent paper by Klauder and Whiting \cite{Klauder3} this formalism
was extended to include additional variables that can be thought of
as gauge degrees of freedom.  In this paper, I consider
first class constraints and their related gauge symmetries.

\section{Coherent State Path Integral}

A generalized coherent state may by defined in the following
way \cite{Perel}.  Let $G$ be a Lie group acting on a Hilbert space.
Let $ \{ |\psi_g \rangle \}$ be a system of states where $|\psi_g \rangle
= U(g) |\psi_0 \rangle ,\  g \in G $.
\ \ $|\psi_0 \rangle $ is a fixed vector from the
Hilbert space (often called the fiducial vector). $ U(g)$ is a unitary
representation of the group $G$ acting on the Hilbert space.
Two states are defined to be equivalent if they differ only by a phase
factor. So if $H$ is the isotropy subgroup in G such for
$h \in H$,

\be
U (h) | \psi_0 \rangle = e^{i \theta (h)} \  |\psi_0 \rangle,
\ee

\noindent
then it is clear from this that each inequivalent state is labeled by a
member of the left coset
space $ G/H$.  For convenience, we shall label the points in this space
by $x \in G / H$ and the coherent state vector
by $ | x \rangle $.
These states do not in general form a
orthonormal basis.  However, they do admit a resolution of unity,

\be
\ID = \int | x \rangle \langle x | \  d \mu (x),
\label{resun}
\ee

\noindent
where $d \mu (x)$ is a positive measure.  These states form an
(over)complete set of states on  the Hilbert space.
We can represent a vector in our Hilbert space as
a function of $x$ by defining the function to be

\be
\psi (x) \equiv \langle x | \psi \rangle.
\ee

\noindent
{}From the resolution of unity \req{resun}, it can be seen that
the inner product on this function space is just the normal
inner product on $L^2$

\be
\langle \psi | \phi \rangle = \int \overline{\psi (x)} \  \phi (x)
\ d \mu (x).
\ee

\noindent
The overlap function $ {\cal K} (x';x) \equiv \langle x' | x \rangle $
is the reproducing kernel on this space.

\bea
\psi (x') & = & \int {\cal K} (x';x)\  \psi (x) \ d \mu (x)
\nonumber \\[1ex]
{\cal K} (x'';x) & = & \int {\cal K} (x'';x')\ {\cal K} (x';x) \ d \mu (x')
\label{last}
\eea

\noindent
These (\ref{resun} - \ref{last}) are the basic ingredients
for a coherent state representation.

Using these basic ingredients, we can construct a path integral
(for more details see \cite{Klauder3} and \cite {Klauder1}).  We start
with the matrix element of the Hamiltonian evolution
\hbox{$\langle x'' | \exp - {i \over \hbar}
\hat H (t''-t') | x' \rangle$}.
By inserting a resolution of unity at each time slice,
we can split the time variable into N pieces and write the total
evolution in terms of the evolution between the time slices.

\vbox{
\bea
 \langle x'' |  e^{ -{i \over \hbar} \hat H(t''-t')}  | x' \rangle
& = & \int d \mu (x_1) \ \langle x'' | e ^ {- { i\over \hbar}
 H (t'' - t' - \varepsilon) } |
x_1 \rangle \langle x_1 | e ^ {- {i \varepsilon \over \hbar} \hat H}
 | x' \rangle
\nonumber \\[1ex]
& \vdots  &  \nonumber \\[1ex]
 \langle x'' |  e^{- {i \over \hbar} \hat H(t''-t')}
 | x' \rangle & = &  \left(
 \int \ldots \int \prod_{n=1}^{N} d \mu (x _{n}) \right)
 \prod_{n=0}^{N} \langle x_{n+1} | e^{- {i \varepsilon \over \hbar}
 \hat H}  | x _{n} \rangle  \nonumber \\[1ex]
x'' = x_{N+1},
\hskip .2in  x'  & = & x_0,
\hskip .2in  \varepsilon = (t'' - t') / (N + 1)
\label{path1}
\eea
}

\noindent
In the limit ($\varepsilon \rightarrow 0$),
if the paths are  continuous and differentiable, then
we can then make the following approximations.

\vbox{
\bea
\langle x_{n+1} | e^{- {i \varepsilon \over \hbar}  H} | x _{n} \rangle
& \approx & \langle x_{n+1} | \ID - {i \varepsilon \over \hbar} \hat H
| x_n \rangle \nonumber \\[1ex]
& \approx & \langle x_{n+1} | x_n \rangle \left(
1- {i \varepsilon \over \hbar}
 { \langle x_{n+1}|  \hat H | x_n \rangle \over
 \langle x_{n+1} | x_n \rangle } \right)
\nonumber \\[1ex]
& \approx & \left( 1- \varepsilon \langle x_{n+1}| \left(
 { | x_{n+1} \rangle - | x_{n} \rangle \over \varepsilon }
 \right) \right) \left(
1- {i \varepsilon \over \hbar} { \langle x_{n+1}|  \hat H | x_n \rangle \over
 \langle x_{n+1} | x_n \rangle } \right)
\nonumber \\[1ex]
& \approx& ( 1- \varepsilon \langle x_{n+1} | {d \over d t}
| x_{n+1} \rangle ) \ ( 1 - {i \varepsilon  \over \hbar} \langle x_{n+1}|
 \hat H | x_{n+1} \rangle )
\label{path2}
\eea
}

\noindent
We define the symbol $H(x) \equiv \langle x | \hat H | x \rangle$.
Then we can re-exponentiate these two terms, keeping terms
up to ${\cal O} (\varepsilon)$, and place them back into
the form above \req{path1}.

\be
\langle x'' |  e^{ -{i \over \hbar} \hat H(t''-t')}  | x' \rangle
= \int \ldots \int \prod_{n=1}^{N} d \mu (x _{n})
\prod_{n=1}^{N+1} e^{- {i \varepsilon \over \hbar} \left(
{i \hbar} \langle x_n | {d \over d t}| x_n \rangle + H(x_n)  \right)}
\label{path9}
\ee

\noindent
In the continuum limit, we have the following
formal expression for the path integral.

\be
\langle x'' |  e^{ {i \over \hbar}  H (t''-t')} | x' \rangle =
\int D x  \exp \left \{ { i \over \hbar} \int \left( i \hbar \langle x |
{d \over d t} | x \rangle  - H (x) \right) d t
\right \}
\label{path3}
\ee

In addition to the symbol $H(x)$ above,
we can also define another symbol for the operator $\hat H$. This symbol
is implicitly defined in terms of the spectral representation of the
operator \cite{KlauderS}.

\be
\hat H = \int h(x) |x \rangle \langle x| d \mu (x)
\label{lowerh}
\ee

\noindent
This symbol $ h(x) $ is called the lower symbol
while $H(x)$ is called the upper symbol.
If such a representation exists and is well defined
(see \cite{Klauder5}) then we have another way to derive the path
integral. Let us consider the
infinitesimal time evolution operator in terms of the lower
symbol.  With the resolution of unity and the
above definition \req{lowerh}, we can write this operator in the
following form.

\be
\ID - {i \varepsilon \over \hbar} \hat H =
\int \left[ 1 - {i \varepsilon \over \hbar} h(x) \right]
|x \rangle \langle x | d \mu (x)
\ee

\noindent
Then by exponentiating both sides, dropping terms of order
$\varepsilon ^2$, and apply repeated operations of
this operator, we can build up a finite time displacement operator.

\be
e^ {- {i \over \hbar} \hat H T } =
\int \ldots \int \prod _{n=1}^N
e^{ - {i \varepsilon \over \hbar} h(x_n) }
| x_n \rangle \langle x_n| d \mu (x_n)
\qquad {\rm where} \qquad T = N \varepsilon
\ee

\noindent
The Hamiltonian evolution matrix then takes the form

\be
\langle x'' | e ^{ -{i \over \hbar} \hat H (t''-t') } | x ' \rangle
= \int \dots \int \prod _{n=1} ^{N+1} \langle x_n| x _{n-1} \rangle
\prod _{n=1} ^N e^{ -{i \varepsilon \over \hbar} h(x_n) } d \mu (x_n).
\label{path10}
\ee

\noindent
If we make similar approximations as we did in \req{path2}, in the
continuum limit, we have the same form of the path integral has in
\req{path3} but with $h(x)$ replacing $H(x)$.

\be
\langle x'' |  e^{ {i \over \hbar}  H (t''-t')} | x' \rangle =
\int D x  \exp \left \{ { i \over \hbar} \int \left( i \hbar \langle x |
{d \over d t} | x \rangle  - h (x) \right) d t
\right \}
\label{path11}
\ee

This second form of the symbol of $\hat H $ is related
to the first form by

\be
H(x) = \int h(x') | \langle x | x' \rangle | ^2 d \mu (x') .
\ee

\noindent
In general these two symbols will not be equivalent.  However
for a suitable choice of coherent states, the difference between
them will only be of order $\hbar$. Therefore, when we take the
stationary phase approximation to the path integral, both $h(x)$
and $H(x)$ will lead to the same equations of motion.

Although there have been attempts to regularize the ordinary
configuration space path integral by introducing additional terms
(see \cite{Klauder2} for a good review) these attempts have
met with limited success.  However the coherent state path integral
is inherently a path integral over the phase space, and because
of this it is possible to regularize with path integral
by changing the measure to a pinned Wiener measure.
This measure originally came from the study of Brownian motion.
The probability density of a particle undergoing Brownian motion
is governed by the diffusion equation.  The equation for the density
$\rho (t'';t')$ at $t''$ starting with initial data at time $t'$
is given by

\be
{\partial \rho (t'';t') \over {\partial t''}} = {1 \over 2} \nu \Delta
\rho (t'';t'),
\ee

\noindent
where $ \Delta $ is the Laplace-Beltrami operator.  For
an example, let the metric be a flat metric ($ d \sigma ^2 = d p ^2
+ d q^2 $), then the fundamental solution of this equation is given by

\be
\rho (t''; t') = {1 \over { 2 \pi \nu (t''- t') }}
\exp \left[ -{(p'' - p')^2 + (q''- q')^2  \over 2 \nu (t''-t')}
\right].
\ee

\noindent
The most important property of this solution for the Wiener measure
is that the density $\rho (t'';t')$ posses
the following product rule \cite{Klauder5}.

\be
\rho (t'''; t') = \int d p'' \ d q'' \ \rho (t''';t'')
\rho (t''; t')
\ee

\noindent
This product rule can be repeated to form a lattice in the time direction.

\bea
&& \rho (t''; t') = \int \left( \prod _{i=1} ^N
d p_i \ d q_i \right)  \left( {1 \over { 2 \pi \nu  \varepsilon }}
\right) ^{N}
\left( \exp { \sum _{i=0} ^N -{(p_{i+1} - p_{i})^2 + (q_{i+1} - q_i)^2
\over 2 \nu \varepsilon}} \right)
\nonumber \\[1ex]
&& (q'',p'') = (q_{N+1}, p_{N+1}),
\hskip .2in
 (q', p')  = (q_0,p_0),
\hskip .2in \varepsilon = (t'' - t') /  N
\eea

\noindent
In the continuum limit, we now have a formal expression for the
Wiener measure which is pinned for both $q$ and $p$ at $t''$ and $t'$.

\be
d \mu _W ^\nu (p,q) = {\cal N} e ^{ -{1 \over 2 \nu} \int \dot p ^2
+ \dot q ^2  \ d t }\  D q \ D p
\ee

\noindent
Writing this in a more general way to include other choices for the
metric, we have

\be
d \mu _W ^\nu (x) =  {\cal N} e ^{-{1 \over 2 \nu} \int
 \left( { d \sigma (x)  \over d t} \right) ^2 d t } \  D x.
\label{wiener}
\ee

\noindent
If it is assumed that on the phase space
no point should be distinguishable from any other point then
the metric $d \sigma (x) ^2$ on the phase space should
be homogeneous.  Therefore, the metric should be chosen such that
the resulting geometry has constant curvature.  Different choices
of the geometry lead to different kinematical variables on which we
quantize the system, for further details see \cite{Klauder2}.
For example, the flat case leads to quantization with the ordinary
Heisenberg pair of operators.  The constant positive curvature
case leads to an underlying quantum kinematical spin
operators $S_i$ where $[S_i,S_j] = i \varepsilon _{ijk} S_k$.

The measure in \req{path3} can now be replaced by the well defined pinned
Wiener measure \req{wiener}. This is done by the addition
of the extra factor

\be
e^{ - {1 \over 2 \nu} \int \left( {d \sigma \over dt} \right)^2 dt}.
\ee

\noindent
In the limit $\nu \rightarrow \infty$, this term vanishes and we
are left with our original path integral. In this limit,
Daubechies and Klauder \cite{Klauder5} showed that the appropriate
symbol to use is the lower symbol $h(x)$.

\be
\langle x'' | e ^{- {i \over \hbar} \hat H T} | x' \rangle
= \lim _{\nu \rightarrow \infty} {\cal N}
\int d \mu _W ^{\nu} (x) \exp -{i \over \hbar} \left\{ \int
i \hbar \langle x | {d \over d t} | x \rangle  + h (x)
 \right\}
\ee

I would now like to consider this formalism with a system of first
class constraints.

\section{Classical Constraints}

To begin with, let us consider a $2M$ dimensional phase space labeled by
coordinates $ y ^a $.  On this phase space, we will consider a
system of $N$ first class constraints given by $\phi_i (y ^a) = 0$.
Let these constraints form a closed algebra with respect to the Poisson
Bracket,

\be
[ \phi_i , \phi_j ] = C_{i j} ^k \ \phi_k.
\ee

\noindent
Let them also be complete. In other words, the constraints also commute with
the total Hamiltonian on the constraint surface such that the time evolution
does not generate further constraints.

\be
[ \phi_i ,  H ] = {d \phi_i \over d t} \approx 0
\label{constr}
\ee

\noindent
Such constraints can always be Abelianized locally by a
canonical transformation \cite{Henn1}.  However, the local coordinate patch
may not cover the entire constraint surface.
For this paper we will assume that we can work on one coordinate
patch. After Abelianization, the constraint equation can now be written as

\be
p_i = 0, \qquad i =1, \ldots , N .
\label{const}
\ee

\noindent
The gauge orbits are then along $q^i$. The reduced phase space can be
labeled by $(2M-2N)$ variables $ z^b$. The Poisson bracket
algebra becomes

\be
[p_i,q^j] = \delta _i^j \qquad [q^i, z^a] = 0
\qquad [p_i, z^a ] = 0 \qquad [ z^a, z^b] = C ^{ab}_c z^c .
\ee

\noindent
The normal coordinates $(p_i,q^j)$
close under the Poisson bracket and commute with the reduced
phase space variables $(z^a)$. We will now map this set of coordinates onto
their related operator such that

\be
[\hat Q^i, \hat P_j ] = i \hbar \delta ^i_j
\qquad [\hat Q^i, \hat Z^a] = 0
\qquad [\hat P_i, \hat Z^b] =0 .
\label{oper1}
\ee

\noindent
The operators $\hat Z^a$ can also be broken into pairs of Heisenberg
operators.  However, we will not need to make use of this other than
the fact that the reduced phase space has its own coherent state
representation.

\be
| z^a \rangle \equiv U(z^a) | \eta \rangle
\ee

\noindent
We will make use of this reduced phase space
coherent state throughout out the paper.

The coherent state on the full phase space may be written as

\be
| y \rangle \equiv |p_i, q^j , z^b \rangle =
e^{if(p,q,z)}  e^{-{ i\over \hbar} q^j \hat P_j}
e^{{i\over \hbar} p_i \hat Q^i}
U (z^b) | \eta' , \eta \rangle .
\label{state0}
\ee

\noindent
where $|\eta' , \eta \rangle = | \eta ' \rangle \otimes
| \eta \rangle$, the direct product of the fiducial vector
for the reduced phase space and the a fiducial vector for the normal
coordinates $(p_i,q^j)$. The first term is just a phase factor. It
appears in the path integral in terms of the symplectic one form
which determines which coordinate will play the roll of the momentum
and position.  This one form can be changed by the addition of a total
derivative in the action (see \cite{Klauder4}).  We will assume that we
are free do this.  The result being that we can write the above
coherent state as the direct product of the reduced phase space
coherent state with the normal coordinates coherent state.

\bea
&& |p_i, q^j , z^b \rangle = | p_i, q^j \rangle
\otimes |z ^a \rangle
\nonumber \\[1ex]
&& |p_i,q^j \rangle = e^{-{ i\over \hbar} q^j \hat P_j}
e^{{i\over \hbar} p_i \hat Q^i} | \eta ' \rangle
\label{state1}
\eea

So we have constructed a coherent state in the new coordinate system.
We would now like to show that this coherent state admits a resolution
of unity.  In the original coordinate system $({y'}^b)$ before we
Abelianized the constraints, the coherent state did admit a resolution
of unity.

\be
(p'_i,{q'}^j) = ({y'}^a) \qquad i,j = 1, \ldots M
\ee

\be
|p'_i,{q'}^j \rangle = e^{ - {i\over \hbar} {q'}^j \hat P'_j}
e^{{i\over \hbar} p'_k \hat {Q'}^k} | \eta \rangle
\ee

\be
\ID = \int |p'_i,{q'}^j \rangle \langle p'_i,{q'}^j|
\prod_{i,j=1} ^M {dp'_i d{q'}^j \over 2 \pi}
\ee

\noindent
The new measure $d \mu(q,p) = d \mu(q',p')$ because the transformation
is canonical and therefore the Jacobian is one.  The resolution of unity
on the new set of coordinates is

\be
\ID = \int |p_i,q^j,z^a \rangle \langle p_i,q^j,z^a|
\left( \prod_{i,j=1} ^N {dp_i dq^j \over 2 \pi} \right) d \mu(z^a)
\qquad i,j = 1, \ldots, N .
\ee

\noindent
We must take care because the transformation may not be one to one.
The above integration is on the whole of the phase space and not on
the image of the original phase space. Now we would like to apply
the constraint to this coherent state.

Because the constraint equation \req{const} is originally a classical
object, there are two ways in which we may apply the constraint.
The first is to apply the constraints to the classical variables $ p_i $.
We will consider this case in this section.  The other way is demand
that the operators $\hat P_i$ on physical states is zero,
similar to Dirac quantization.  We will
consider this  approach in the next section.

The classically constrained coherent state ($p_i = 0 $) becomes

\be
|q^i, z^b \rangle \equiv | p_i =0 ,q^j, z ^b \rangle =
e^{-{i \over \hbar} q^j \hat P_j} | \eta ' \rangle
\otimes | z ^a \rangle .
\ee

\noindent
We want to choose a fiducial vector such that
 $\langle \eta |\hat P_i| \eta \rangle =
\langle \eta | \hat Q_i | \eta \rangle = 0 \ \forall i$.
Such a vector is called physically centered, and
can alway be found \cite{KlauderS}.
Now, looking at the expectation values of $\hat P _i$, we see
that the constraint becomes fuzzy (higher orders terms of the
constraint operators are not zero).

\vbox{
\bea
\langle  q^i, z ^a |\hat P_i | q^i, z ^a \rangle & = &
\langle \eta | \hat P_i | \eta \rangle = 0
\nonumber \\[1ex]
\langle  q^i, z ^a |\hat P_i^2 | q^i, z ^a \rangle & = &
\langle \eta | \hat P_i^2 | \eta \rangle = {\hbar \over 2}
\nonumber \\[1ex]
& \vdots  &  \nonumber \\[1ex]
\langle q^i, z ^a | \hat P_i ^n | q^i, z ^a \rangle & = &
\langle \eta | \hat P_i^n | \eta \rangle = {\cal O} (\hbar)
\label{fuzz}
\eea
}

\noindent
The classical constraints can be understood as fixing the center
of a wave packet instead of forcing the wave function to collapse into
an eigenstate of the constraint operators.

We have seen how to construct states that classically satisfy the
first class constraints.  Next, let us consider how we can construct
the path integral using these states.
We can use the resolution of unity on the full phase space to construct
the path integral.  Then at each time step in \req{path1} or \req{path10},
we will force the constraint equation to be obeyed by projecting onto
$p_i=0$ with the standard delta function.  Because we are only dealing
with first class constraints, we do not have to include another term such as
a determinate for second class constraints in our path integral.

\be
\delta (p_i) = \int_{- \infty}^{\infty}
d \lambda^i  \  e^ {-i \lambda^i p_i}.
\label{project}
\ee

\be
 \langle y'' |  e^{i \hat H(t''-t')} | y' \rangle =
 \int \ldots \int \prod_{n=1}^{N} d \mu (y _{n}) \ d \lambda^i_{n}
 \prod_{n=0}^{N} \langle y_{n+1} |
 e^{-{ i \varepsilon \over \hbar}  \hat H}
 e^ {-i \lambda^i_{n} {p_i}_n} | y _{n} \rangle
\label{path7}
\ee

\noindent
Now we rescale $\lambda$ by $\hbar / \varepsilon$
and adjust the measure accordingly.

\vbox{
\bea
&& \langle y'' | e^{i \hat H (t''-t')} | y' \rangle =
\int \ldots \int \prod_{n=1}^{N} d \mu (y_{n}) \  d  \lambda^i_{n}
\ \prod_{n=0}^{N} \langle y_{n+1} | e^{-{i \varepsilon \over \hbar}
 (\hat H +  \lambda^i_n {p_i}_n )} | y _{n} \rangle
\nonumber \\[1ex]
&& y_0= y' = (p =0 ,q', z')  \hskip .5in  y_{N+1}= y'' =
 (p =0 ,q'', z'')
\label{path4}
\eea
}

\noindent
In the continuum limit, the path integral becomes

\bea
&&
\int D y  D \lambda  \exp - {i \over \hbar} \left[ \int
i \hbar \langle y | {d \over d t} | y \rangle  + H_T (y) \right]
\nonumber \\[1ex]
&& {\rm where} \qquad  H_T (y) = H(y) + \lambda ^i p_i .
\label{symb}
\eea

\noindent
With a physically centered fiducial vector, the symbol $H_T(y)$
can be defined as  \hbox{ $ \langle y |\hat H (y) + \lambda ^i \hat P_i
| y \rangle $} which is equal to symbol above \req{symb}.
Because the Lagrange multiper term is linear in $p_i$, this term
is equivalent when we switch between upper and lower symbols. So
the lower symbol for the total Hamiltonian is
\hbox{ $ h(y) +  \lambda(y) ^i p_i(y)$}.

At this point, we have derived the path integral for the total
Hamiltonian. If we apply a stationary phase approximation,
we get the normal classical equation of motion
plus the constraint equations $p_i = 0$. However, we would like to
change to the pinned Wiener measure in the above path integral.
In addition we would like to include the extended variables
$\lambda^i$ into this measure.
We would also like to find the reduced Hamiltonian and its
associated path integral.  In so doing, we must take care to choose a
suitable metric for the Wiener measure.

To find the reduced phase space Hamiltonian and the associated
path integral,
let us return to the lattice path integral \req{path4}.  Before
taking the continuum limit, we want the states at each time slice
to satisfy the constraint equation ($p_i=0$). Integrate along
$\lambda ^i_n$ leads to a $ \delta ({p_i}_n) $ at time step.

\be
 \langle y'' | e^{-{i \over \hbar} \hat H (t''-t')} | y' \rangle =
 \int \ldots \int \prod_{n=1}^{N} d y_{n} \  \delta ({p_i}_n)
 \ \prod_{n=0}^{N} \langle y_{n+1} | e^{-{ i \varepsilon \over \hbar}
 \hat H} | y _{n} \rangle
\ee

\noindent
Now, because we have removed the extended variables, we can take the
continuum limit and change the measure to the Wiener measure.

\be
 \langle y'' | e^{-{i \over \hbar} \hat H (t''-t')} | y' \rangle =
 \int d \mu _W ^{\nu} (y) \ \delta [p_i]
 \exp -{i \over \hbar} \left\{ \int
 i \hbar \langle y | {d \over d t} | y \rangle  + h (y)
 \right\}
\ee

If we had just transformed the labeling space of the coherent state
by defining $|\tilde y \rangle \equiv | y \rangle$ as in
\cite{Klauder8}.  Then we would have to carry over the metric for the
Wiener measure \req{wiener} from the
original phase space to the new coordinates.  Klauder calls this a
shadow metric.  However we have switched our coherent state to be defined
on the new phase space, and we have replaced the operators with the new
operators \req{oper1}.  Because of this, it is not clear which metric should
be placed on this new phase space.  The metric should
however be compatible with  our constraints.

We wish to integrate the path integral in the $p_i$ direction to
remove the delta function and fix $p_i=0$.
Because we do not want to introduce any extraneous
coupling between the reduced phase space coordinates ($z$) and the normal
coordinates ($p,q$) let us choose a metric on the phase space that can be
separated,

\be
d \sigma (y)^2 = d  \sigma (z)^2 + d \sigma (p,q)^2.
\ee

\noindent
In addition, we want the metric $ d \sigma (p,q)^2$
to be consistent with the gauge transforms
($ q^i \rightarrow q^i + f^i$).
Therefore, the metric should be independent of $q^i$.
Also, the metric should be well defined for $ p_i=0 $.
For example, in two dimensions, we can choose the metric to
have the form

\be
d  \sigma (p,q)^2 = g_{11} (p) d p^2 + 2 g_{12} (p)
d p d q + g_{22} (p) d q^2.
\ee

\noindent
Now, we  can integrate over the $p$. The delta function will fix $p=0$
for each time slice.  So the metric will become

\be
d \sigma (q) ^2 = g_{22} dq^2 ,
\ee

\noindent
where $g_{22}$ is now just a constant.

Now, we can also integrate along the gauge orbits $q^i$.  Because the integral
is regularized, on first appearances,
we do not have to gauge fix these orbits to remove the
infinity redundancies.  Because $g_{22}$ is just a constant on the
constraint surface, we can rescale $\nu$ and/or $q^i$, such that the
metric is just $dq^2$.  The measure for our Wiener measure should then take
the form

\be
 d \mu^{\nu}_{W} (z , q) =
 {\cal N} e ^{-{1 \over 2 \nu} \int
 \left( { d \sigma (z)  \over d t} \right)^2 +
 \left( { dq  \over d t} \right)^2 }
 D z D q .
\label{measn}
\ee

Turning to the terms in the exponent in the path integral,
the term $ \langle y | {d \over dt} | y \rangle $ can be
written in the following way
using the definition \req{state1}.

\be
\langle y | {d \over dt} | y \rangle = - {i \over \hbar}
 p_j {d q^j \over dt}
+ \langle z | {d \over dt} | z \rangle
\ee

\noindent
With $p_i = 0$ the first term drops out, and we can write the
path integral in the continuum limit as

\be
\int d \mu^{\nu}_{W} (z , q)
\exp \left \{ - {i\over \hbar}  \int \left( i \hbar \langle  z|
{d \over d t} |  z \rangle  + h(z,q) \right)
d t \right \} .
\label{classpath}
\ee

\noindent
The symbol $h(q,z)$ is defined from the full phase space symbol
$h(p,q,z)$ with $p_i$ set equal to zero.  All though there is no
$p_i$ dependence in the symbol, there may still be $q^i$
dependence.  We will
discuss this point in a bit.  For now, let us try to extend the
Wiener measure to the extended phase space.

Let us go back to our earlier definition of the path integral \req{symb},
and instead of fixing the states at each time slice,
we will work on the extended phase space and its associated
total Hamiltonian. In so doing, we should
regularize the new measure which includes the extended coordinates.
Because of the of the above construction, it is natural to choose the metric
on this extended phase space as

\be
d \sigma (y, \lambda)^2 = d  \sigma (z)^2 + d \sigma (p,q)^2 +
d \lambda^2.
\ee

\noindent
The Wiener measure now takes the form

\be
 d \mu^{\nu}_{W} (y , \lambda) =
 {\cal N} e ^{-{1 \over 2 \nu} \int
 \left( { d \sigma (z)  \over d t} \right) ^2 +
 \left( { d \sigma (p,q) \over d t} \right) ^2 +
 \left( { d \lambda  \over d t } \right) ^2  d t}
 \ D z  D p  D q D \lambda ,
\ee

\noindent
and the path integral \req{path4} becomes

\be
\int d \mu^{\nu}_{W} (y , \lambda)
\exp \left \{- {i \over \hbar} \int \left( i \hbar \langle y |
{d \over d t} | y \rangle  + h (y)
+  \lambda^i p_i  \right) d t \right \}.
\ee

We can now integrate $\lambda$ with our definition above.  Let us
work out an example with only one constraint and a flat metric on
the constraint variables
$d \sigma (p,q)^2 =  d p^2  +  d q^2 $.
The only terms involving $\lambda$ are

\bea
 \int d \mu^{\nu}_{W} (\lambda) \ \exp ( i \lambda p) & = &
 {\cal N} \int  \prod _{n=1} ^N
d \lambda _n   \left( {1 \over { 2 \pi \nu  \varepsilon }}
\right) ^{N \over 2}
\nonumber \\[1ex]
&&  \exp \left( { \sum _{n=0} ^N -{(\lambda _{n+1} - \lambda _n)^2
\over 2 \nu \varepsilon}}  -{ i \over \hbar}  \lambda _n p_n \right) .
\eea

\noindent
Integrating over one of the $\lambda_n$'s,

\bea
&& \int_{- \infty} ^ {\infty}  d \lambda _n
\left({1 \over { 2 \pi \nu  \varepsilon }} \right) ^{1 \over 2}
\exp \left[  - {1 \over 2 \nu \varepsilon} \left( (\lambda _{n+1} -
 \lambda _n)^2 +
(\lambda _{n} - \lambda _{n-1})^2  \right)
-{ i \over \hbar}  \lambda _n p_n  \right]
\nonumber \\[1ex]
&& =  {1 \over \sqrt{2} } \exp \left[ -{1 \over 2 \nu \varepsilon}
\left( {1 \over 2} (\lambda_{n+1} - \lambda_{n-1})^2
+ {i \over \hbar} \varepsilon  \nu p_n
(\lambda_{n+1} + \lambda_{n-1}) + { \varepsilon^2
\nu ^2 \over 2 \hbar^2}  p_n^2 \right)
\right] .
\eea

\noindent
If we were to take the limit $\nu \rightarrow \infty$
at this stage, the last term  ($p_n^2$) dominates, and
we see that the delta function is still hidden in this expression.

\be
\lim _{\nu \rightarrow \infty} \exp - \nu  p_n^2
=  \sqrt{\pi \over \nu }\  \delta (p_n).
\ee

\noindent
However, we should consider taking the limit of all the $\mu$
together.  So let us
continue to integrate over all possible $\lambda_n$'s without taking
the limit yet.  In so doing

\bea
 && \int d \mu^{\nu}_{W} (\lambda)  \exp ( -{i\over \hbar} \lambda p)  =
 {\cal N} \left( {1 \over 2 \pi (t_{N+1} - t_0) } \right) ^{1 \over 2}
 \exp \left[ - {(\lambda _{N+1}
 - \lambda_0 )^2  \over 2 \nu (t_{N+1} - t_0) } \right]
 \nonumber \\[1ex]
 && \qquad \exp \left[ {1 \over (t_{N+1} - t_0)} \left[ \left( \sum
_{j = 1} ^{N}
 - {i \over \hbar} p_j (t_j - t_0) \right) \lambda _{N+1}
+ \left( \sum _{j=0} ^{N-1}
 - {i \over \hbar} p_j (t_{N+1} - t_j) \right) \lambda _0 \right] \right]
\nonumber \\[1ex]
 && \qquad \exp \left[ - { \nu \over \hbar ^2
 (t_{N+1} - t_0)} \sum _{j,k = 1} ^{N-1}
 p_j p_k (t_{N+1} - t_j) (t _k - t_0) \right].
\eea

\noindent
In the continuum limit, this becomes

\bea
 && \int d \mu^{\nu}_{W} (\lambda) \exp ( -{i \over \hbar} \lambda p)  =
 {\cal N} \left( {1 \over 2 \pi (t_{N+1} - t_0) } \right) ^{1 \over 2}
 \exp \left[ - {(\lambda _{N+1}
 - \lambda_0 )^2  \over 2 \nu (t_{N+1} - t_0) } \right]
 \nonumber \\[1ex]
 && \qquad \exp \left[ {1 \over (t_{N+1} - t_0)} \left( \int
 -{i \over \hbar} p(t) (t - t_0) dt \right) \lambda _{N+1} + \left( \int
 -{i \over \hbar} p(t) (t_{N+1} - t) dt \right) \lambda _0 \right]
\nonumber \\[1ex]
 && \qquad \exp \left[ - { \nu \over \hbar^2 (t_{N+1} - t_0)}
 \int_{t_0} ^{t_{N+1}}  \int_{t_0} ^{t}  p(t) p(s)
 (t_{N+1} - t) (s - t_0) ds\ dt \right].
\label{conlim}
\eea

Because $\lambda$ is just a Lagrange multiper, after integrating, the final
answer should not be dependent on its initial or final value.
The first term in the limit $\nu \rightarrow \infty$ is zero
if $ \lambda_N+1 \neq \lambda_0 $. This forces $\lambda_N+1 =
\lambda_0$.  This would seem to make it impossible to fold together
two propagators to make a propagator over a longer time.

\be
\langle y'''| e ^{-{i \over \hbar} \hat H (t'''-t')} | y ' \rangle
= \int D y'' D \lambda \
\langle y'''| e ^{-{i \over \hbar} \hat H (t'''-t'')} | y '' \rangle
\langle y''| e ^{-{i \over \hbar} \hat H (t''-t')} | y ' \rangle
\ee

\noindent
But this condition is forced on us at the last step of the integration
and does not need to be set before hand. So we can fold together
two propagators and as a last step let $\lambda_{N+1} = \lambda_0$.

The second term in \req{conlim} gives us a possible phase factor.

\be
\exp \Big[ i \lambda \int p(t) dt \Big] =
\exp \Big[ i \lambda  p_{\rm ave}   (t'' - t')  \Big]
\label{phase1}
\ee

\noindent
The last term can be expanded in terms of a saddle point
approximation as $\nu$ becomes large.

\be
 S[p]   =  \int \int \Big( p(t) p(s)
\theta (s -t)  (t_{N+1} - t) (s - t_0) \Big)  ds \  dt
\ee
\bea
\Rightarrow \qquad {\delta S[p]  \over \delta p } =  p (s) \int \Big(
 \theta (s - t)  (t_b -t)(s - t_a) + \theta (t - s)
(t_b -s ) (t - t_b)  \Big) dt = 0
\nonumber \\[1ex]
\eea

\noindent
This integrand is always positive.  Therefore,
for $\delta p$ arbitrary, this implies that $ p(t) = 0 $.
Once again, we get back to our original delta function.
With this delta function, as we integrate our path integral along $p$, we
see that the second term \req{phase1} also drops out.
The extended phase space, with the above choice for the metric
on the phase space leads to an equivalent path integral to the one
derived before \req{classpath}.

Either integrating over each time slice or choosing to extend our
phase space, the resulting ``classical''
reduced Hamiltonian symbol $H_{cr}$ takes the form

\bea
H_{cr} & \equiv & \langle p,q,z | H(\hat P_i, \hat Q^i, \hat Z^a)
| p,q,z \rangle \Big| _{p_i =0} \nonumber \\[1ex]
& = & \langle z| \otimes \langle \eta  |
 H(\hat P_i, \hat Q^i + q^i \ID, \hat Z^a  )
| \eta \rangle \otimes | z \rangle.
\label{classmb}
\eea

\noindent
Note, if $H(\hat P_i, \hat Q^i, \hat Z^a)$ contains terms like
$F(\hat Q) \hat P ^2$, such a term, for example, might occur in the kinematic
term on a curved surface, then

\bea
\tilde H & = & \langle q, z | F(\hat Q) \hat P^2 | q, z \rangle
\nonumber \\[1ex]
& = & \langle \eta | F (\hat Q + q \ID ) \hat P^2 | \eta \rangle
\nonumber \\[1ex]
& = & \langle \eta | \left( F(q) + {\partial F \over \partial Q}
 \bigg|
_{Q=q} \hat Q  + \ldots \right) \hat P^2 | \eta \rangle
\nonumber \\[1ex]
& = & F(q) \langle \eta | - {\hbar \over 2} ( a - a^{\dagger} )^2
| \eta \rangle + {\cal O} (\hbar^2)
\nonumber \\[1ex]
& = &{\hbar \over 2} F(q) + {\cal O} (\hbar^2).
\eea

\noindent
The classically reduced Hamiltonian symbol
$H_{cr}$ is still in general dependent on the
$q^i$ the gauge orbits.  However such a term can only appear for
terms of order $\hbar$ or higher.

\bea
H_{cr}(z,q) & = & \langle z | H( \hat Z^a) |z \rangle + \hbar H_1 (z,q)
\nonumber \\[1ex]
 & =& H_0 (z) + \hbar H_1(z,q),
\label{symb2}
\eea

\noindent
where $H_0$ is the classical Hamiltonian on the
reduced phase space.  Even if there is no dependence on $q^i$ in
$H_1$, such a term will not, for most reasonable Hamiltonians, be zero.
Likewise for the lower symbol.

\bea
\tilde H ' & = & \int p^2 f(q) |p,q \rangle \langle p,q | dp dq
\nonumber \\[1ex]
& = & \hat P^2 F(\hat Q) + {\cal O} (\hbar)
\eea

\noindent
If such a term should appear, it would break the original gauge
symmetry.  Because this gauge breaking term(s) will be of the order
$\hbar$, it will not appear in the equations of motion.  However,
the breaking of the gauge symmetry will lead to a classical observable
effect.  This implies that we have not correctly chosen the Hamiltonian
operator for this system.  If the classical Hamiltonian is a polynomial
in $p$ and $q$, then we can clearly redefine the Hamiltonian operator
in terms of the above spectral representation without any problems.
In so doing, we are left with
only the reduced phase space Hamiltonian $h_0(z)$.

At this point, we must still integrate along the
gauge orbits in general. So our path integral is

\bea
&&
{\cal N} \int d \mu^{\nu}_{W} (z , q)
\exp \left \{ -{i \over \hbar} \int  i \hbar \langle z|
{d \over d t} |  z \rangle  + h_0 (z) d t \right \}
\nonumber \\[1ex]
&& d \mu^{\nu}_{W} (z , q) =
{\cal N} e ^{-{1 \over 2 \nu} \int
\left( { d \sigma (z) ^2 \over d t} \right) +
\left( { dq ^2 \over d t} \right) dt}
D z D q
\eea

\noindent
Now that we have removed all the gauge dependencies.  We can integrate
along gauge orbits.  Because the Wiener measure gives us a finite
volume for this integration, we do not have to gauge fix this
path integral.  Instead, we can just integrate the volume and
absorb it into the normalization constant.
This leaves us with the reduced phase space path integral

\bea
\langle y'' | e ^{ - {i \over \hbar} \hat H_T (t''-t')} | y ' \rangle
& = & {\cal N}' \int d \mu^{\nu}_{W} (z)
\exp \left \{ -{i \over \hbar} \int i \hbar \langle z|
{d \over d t} |  z \rangle   + h_0 (z) d t \right \}
\nonumber \\[1ex]
& = & \langle z '' | e ^{ - {i \over \hbar} \hat H_0 (t''-t')}
| z ' \rangle
\eea

\section{Quantum Constraints}

Another approach to applying the constraints is to define the physical
states as being annihilated by the constraint operator
as in Dirac quantization.  For our constraint
($p_i =0$), there is no problem with factor ordering.  So we
can define the physical states as

\be
\hat P_i |\psi _{phys} \rangle = 0 .
\label{const2}
\ee

\noindent
These constraints are sharp, unlike the earlier classical constraint
\req{fuzz}.  This is because we are now forcing the state to collapse
into a momentum eigenstate.

The physical state is therefore the zero momentum eigenstate and the
reduced phase space coherent state,

\be
|q,p,z \rangle _{phys} = |p_i =0 \rangle \otimes |z^a \rangle .
\ee

\noindent
Let us see what happens when we try to write this zero momentum
eigenstate in terms of the earlier coherent state form \req{state1}.

\bea
\hat P_j | q^i, p_j, z^a \rangle _{phys} & = &
\hat P_j \left( e^{-{ i\over \hbar} q^j \hat P_j}
e^{{i\over \hbar} p_i \hat Q^i} | \eta ' \rangle
\right) \otimes | z^a \rangle
\nonumber \\[1ex]
& = & \left( p_j e^{-{ i\over \hbar} q^j \hat P_j}
 e^{{i\over \hbar} p_i \hat Q^i}  | \eta ' \rangle
 +  e^{-{ i\over \hbar} q^j \hat P_j} e^{{i\over \hbar}
 p_i \hat Q^i} \hat P_j | \eta ' \rangle  \right)
\otimes | z^a \rangle
\nonumber \\[1ex]
&= & 0
\nonumber \\[1ex]
\Rightarrow \qquad
 \hat P_j | \eta ' \rangle & = & - p_j | \eta ' \rangle
\eea

\noindent
So the fiducial vector must also be an eigenstate of the momentum
operator.  Continuing, we can further simplify by changing the
order of the terms in the coherent state.  Note also that the
position operator generates translations in the momentum space.

\bea
| q^i, p_j, z^a \rangle _{phys} & = & e^{ - {i \over \hbar} q_i p^i}
e^{ {i \over \hbar} p_j \hat Q^j}
e^{ - {i \over \hbar} q^k \hat P_k} | - p_j \rangle
\otimes | z^a \rangle
\nonumber \\[1ex]
& = & e^{{i \over \hbar} p_j \hat Q^j} | - p_j \rangle
\otimes | z^a \rangle
\nonumber \\[1ex]
& = & | - p_j + p_j \rangle \quad = \quad  |p=0 \rangle
\eea

\noindent
So we have come full circle and back to the zero momentum eigenstate.

These constraint equations also can be solved easily in the Shr\"odingder
representation.  We keep the same ordering of the coherent state
as defined in \req{state1}.

\bea
&& \hat Q^i = x^i \hskip 1in \hat P_i = - i \partial / \partial x^i
\nonumber \\[1ex]
&& \langle x^i |q , p \rangle = \psi (x^i) =
 e^ {-i q^i \hat P_i} e^{i p^i  \hat Q_i}
\phi (x^i) = e^{i p_i q^i} e^{i p_i x^i} \phi (x^i-q^i)
\nonumber \\[1ex]
&& \int d^N x \  \overline{\phi(x^i)} \phi (x^i) = 1
\eea

\noindent
The quantum constraint equation \req{const2} is then solved by
\be
-i {\partial \over \partial x^i} \  \psi (x^i) = 0
\hskip .25in \Longrightarrow \hskip .25in \psi (x^i) = const.
\ee

\noindent
Both the $q^i$ and $p_i$ dependencies drop out of the coherent
state.  We are left with a coherent state that is only dependent
on the reduced phase space coordinates. However, this reduced phase
space coherent is no longer normalizable on the full phase space.
To deal with this, we can let $x$ be restricted to a finite box.
Then let the box go to infinity.  The normalized physical wave function is

\be
\langle x| q,p,z \rangle_{phys} =
 {1 \over \sqrt{ Vol}} \langle x  | z \rangle
\ee

\be
\Rightarrow \qquad |p,q,z \rangle _{phys} \approx
| z \rangle .
\ee

\noindent
This leads us to  using the reduced phase space coherent
state and its resolution of unity to construct  path integral
that only involves the reduced Hamiltonian.  However we will see
that we can still use the full space time to construct a path integral.

The resolution of unity on the full phase space can be shown to
preserve the form of the physical state.

\be
| p = 0 \rangle = \int |p',q' \rangle \langle p',q' | p = 0 \rangle
{d p' d q' \over 2 \pi}
\ee

\noindent
We can also solve part of the Hamiltonian metric element with out
resorting to the using the path integral.  Because the physical
state is the zero eigenstate of momentum, all terms with $\hat P$
drop out.  Because we are assuming only first class constraints
this means that all terms with $\hat Q$ also drop out.  So we are
left with the reduced Hamiltonian.

\bea
\langle p',q',z' | e^{ -{i \over \hbar} H(\hat P, \hat Q, \hat Z) T}
| p,q,z \rangle _{phys} & = &
\langle z' | \otimes \langle p=0 |
 e^{ -{i \over \hbar} H(\hat P, \hat Q, \hat Z) T}
| p = 0 \rangle \otimes | z \rangle
\nonumber \\[1ex]
& = & \langle z' | e^{ -{i \over \hbar}  H_0 (\hat Z) T} | z \rangle
\eea

\noindent
Because the resolution of unity preserves the physical state and the
reduced Hamiltonian does not act upon the normal coordinates, we can
write the path integral on the full phase space.  There are no time
dependencies in the normal coordinates so the first term also only
depends on the reduced coordinates.

\be
\langle y'' | e ^{ - {i \over \hbar} \hat H_T (t''-t')} | y ' \rangle
_{phys}  =  {\cal N} \int d \mu^{\nu}_{W} (p,q,z)
\exp \left \{ -{i \over \hbar} \int i \hbar \langle z|
{d \over d t} |  z \rangle   + h_0 (z) d t \right \}
\nonumber \\[1ex]
\ee

\noindent
If the assumptions that the Wiener measure does not involve cross
terms between $z$ and $p$ and $q$, the integration over $(p,q)$
in the above path integral only give us a volume and can be absorbed into
the normalization.  Once again we arrive at the reduce phase space
path integral.

\be
\langle y'' | e ^{ - {i \over \hbar} \hat H_T (t''-t')} | y ' \rangle
 =  {\cal N}' \int d \mu^{\nu}_{W} (z)
\exp \left \{ -{i \over \hbar} \int i \hbar \langle z|
{d \over d t} |  z \rangle   + h_0 (z) d t \right \}
\nonumber \\[1ex]
\ee

\section{Discussion}

Either using the classical constraints or the quantum constraints,
we are lead to the reduced phase space path integral.
The above derivations relied on the fact that we can find a set
of coordinates for which the constraint equation become simple ($p=0$).
It is easy to see that the classical constraint approach can
easily be carried over to constraints that are not just linear in
the momentum but may depend on higher orders of $p$ and $q$.  The
momentum symbol in the path integral would just be replaced by
the symbol for the constraint.
Ordinarily it is difficult to repeat the  above solution for a quantum
constraint that is not just linear in the momentum. In this case,
it is not clear
which ordering of the operator to use.  Also it may not be possible to
solve the equation for the physical states.  However, Klauder in a recent
pre-print \cite{KlauderP} has constructed a operator version of the
projection operator \req{project}.  This projection operator ties together
these two approaches.

The fact that we arrive at the reduced phase space path integral should
not be surprising with the assumptions we have made.  These assumptions
lead to a simple topology on the phase space.
The difference between Dirac quantization and
reduced phase space tends to appear only with non-trivial topology
(see for examples \cite{dirac}).
It should be possible to extend this ideal to work on a set of
coordinate patches, then we can use constraint coherent states no study some
examples where the reduced phase space and Dirac quantization do
not agree.  In these cases a difference between the classical constraints
and the quantum constraints may appear.  If these two approaches
do not agree then the
coherent states would give a direct relationship between Dirac and
reduced phase quantization.

\section{Acknowledgments}

I would like to thank Steve Carlip for all of his support and time.
I would also like to thank John Klauder for his comments.
This work was supported by National Science Foundation grant PHY-93-57203
and Department of Energy grant DE-FG03-91ER40674.

\end{document}